\documentclass[12pt]{article}
\usepackage{epsfig,cite}
\newcommand{\ben}{\begin{eqnarray}\displaystyle}
\newcommand{\een}{\end{eqnarray}}
\newcommand{\refb}[1]{(\ref{#1})}
\newcommand{\p}{\partial}

\begin{document}

\thispagestyle{empty}

\begin{flushright}
hep-th/0204204\\
MRI-P-020402
\end{flushright}

\baselineskip=18pt

\vskip 3.5cm

\begin{center}
{\Large \bf On the Cosmological Relevance of the Tachyon}\\

\vspace*{6.0ex}

{\large \rm Debajyoti Choudhury,
Debashis Ghoshal,\\ 
Dileep P.\ Jatkar
and Sudhakar Panda\footnote{{\tt 
debchou, ghoshal, dileep, panda@mri.ernet.in}}}

\vspace*{1.5ex}

{\large \it Harish-Chandra Research Institute}\\ 
{\large \it  Chhatnag Road, Jhusi}\\
{\large\it Allahabad 211019, India}

\vspace*{6.5ex}

{\bf Abstract}

\begin{quote}
We analyse of the effective action of the tachyon 
field on a D-brane, of both bosonic as well as superstring theory. 
We find that the non-standard kinetic term of the tachyon field
requires a correction to the Born-Infeld type Lagrangian. The 
cosmological significance of the resulting dynamics is explored.
We also examine if the rolling tachyon can provide an effective
cosmological constant and contrast its behaviour with quintessence.
\end{quote}
\end{center}

\newpage

\setcounter{page}{1}



\section{Introduction}
The perturbative spectrum of open (super-)string around a 
(non-supersymmetric) D-brane contains a scalar which is tachyonic. 
The dynamics of this tachyon field has many unusual features. Sen 
conjectured\cite{origin} that the potential of this field has a 
minimum and when the tachyon condenses to its minimum, 
{\em all} open string excitations become infinitely heavy and hence 
inaccessible to the low energy observer. Of course, for such a thing 
to happen, the Lagrangian describing the tachyon dynamics ought to be 
of an unconventional form. In particular, the effective open string 
coupling is inversely proportional to the value of the tachyon 
potential\cite{as}. Since the potential vanishes at the 
minimum, the open string theory there is infinitely strongly coupled.

More recently, Sen\cite{senroll} has pointed out that the time 
evolution of the tachyon field may have cosmological significance. 
Several authors have initiated an exploration of this 
idea\cite{senmatt,Gibb,FaTy,mukoh,senFT,paddy}. (For other works 
on the cosmological relevance of the tachyon, see \cite{early,early1}.) 
Thanks to unconventional 
form of the tachyon action, cosmology with tachyon matter can lead 
to results different from those obtained with a normal scalar field. 
For example, inflation driven by the tachyon differs from that driven by a 
conventional inflaton. It is also possible that tachyon matter could 
be a candidate for cold dark matter\cite{senroll}\ as, at the minimum 
of its potential, it becomes a pressureless gas. Padmanabhan has 
recently emphasized\cite{paddy}\ that it is always possible to 
construct a potential that leads to a given inflationary scenario. 
Therefore, it is all the more important to study cosmological 
consequences of potentials arising out of a fundamental theory such as 
string theory. With this motivation, we study inflation driven by 
the rolling of the tachyonic scalar on a (non-supersymmetric) 
D3-brane. 

At this stage we should point out the limitations of the scope of our
study. We assume, following Ref.\cite{Gibb}, that the open string
tachyon couples minimally to the graviton, but to {\em no} other closed
string mode. This is rather a drastic assumption since there is no
reason to ignore the coupling to, say, the dilaton. Moreover
it is not known how to stabilize some of the other closed string moduli, 
such as the volume of compactification, the dynamics of which might
affect that of the tachyon. We will, however, take the pragmatic
point of view that some as yet unknown mechanism freezes these
unwanted moduli and leaves us with an effective theory of the tachyon
minimally coupled to Einstein-Hilbert gravity in four dimensions, 
as proposed by Gibbons\cite{Gibb}. 

The plan of the paper is as follows.
In the next section, we review the relevant aspects of tachyon 
dynamics following Refs.\cite{Gibb,FaTy}. In Sec.~3, based on a 
comparison between the Born-Infeld action\cite{as,effective} and 
the tachyon effective action\cite{GeSh,KuMaMoB,KuMaMoS} obtained 
in the so called background independent string field theory 
(B-SFT)\cite{bsft}, we propose a correction to the BI action used 
in the recent works\cite{senroll,senmatt,Gibb,FaTy,mukoh,senFT,paddy}.
The consequences of the extra term are worked out in Secs.~4 and 5
for bosonic and superstring respectively. We find that the tachyon
does not give enough inflation. Sec.~6 is an analysis of 
the slow roll conditions in these models. Perhaps not unexpectedly, 
these conditions are not met. In the 
concluding section, we examine whether the rolling tachyon could 
be an alternative to cosmological constant or quintessence.

\bigskip

\noindent{\bf Note Added:} Since an earlier version of our paper,
a number of authors have studied various aspects related to tachyon
driven inflation. Most notably Ref.\cite{koflin}, (see also
\cite{ShW}), argued on general grounds that inflation driven by
tachyon cannot produce enough expansion. When we re-examined our
results, it was found that incompatible conventions ($2\pi\ell_s^2=1$ 
and $\kappa=1$) were inadvertently used. We correct this unfortunate 
error in the present version. The conclusions are modified in the
light of the new result. 



\section{Dynamics of tachyon matter}
The effective field theory on a (non-supersymmetric) D-brane is 
described by the Born-Infeld Lagrangian
\begin{equation}\label{BornInf}
{\cal L}_{BI} = - V(T)\sqrt{-\det\left[ g_{\mu\nu} + 2\pi\ell_s^2
\left(F_{\mu\nu} + \p_\mu Y^i\p_\nu Y^i + f(T)\p_\mu T\p_\nu T
\right)\right] },
\end{equation} 
where, $1/2\pi\ell_s^2$ is the string tension, $g$ the induced 
metric, $F$ the gauge field strength 
and $Y^i$ are scalar fields describing the transverse motion of 
the brane. Moreover, $T$ is the tachyonic mode on the D$m$-brane,
$V(T)$ the potential of $T$ and $f(T)$ is a function of the 
tachyon. The particular dependence on the potential is a characteristic of 
D-brane physics\cite{as}\ compatible with the conjectures of 
Sen\cite{origin}. The tachyon potential thus plays the role 
of the (inverse) effective coupling of this theory. That the 
derivatives of the tachyon field appears under the square-root 
was pointed out in \cite{effective}\ based on the T-duality 
symmetry of string theory. The arguments in \cite{effective} 
allow for a function $f(T)$ accompanying these derivative terms. 
We shall see this term is non-trivial as the kinetic term of the
effective action following from string field theory does not 
have a canonical form.

Let us couple this system to gravity\cite{Gibb} and neglect 
all fields other than the tachyon. Since our objective is to 
study the cosmological significance of tachyon dynamics, we 
shall assume a Friedmann-Robertson-Walker metric  
\begin{equation}\label{FRW}
ds^2 = - dt^2 + a^2(t)\,\left({dr^2\over 1-kr^2} + r^2 
d\Omega_{m-1}^2\right),
\end{equation} 
as well as a spatially homogeneous and isotropic tachyon field
$T$ which depends only on time. Let us specialize to the case
$m=3$ from now on. The dynamics of this system
is given by the Lagrangian 
\begin{equation}\label{gravT}
\sqrt{-\det g}\,\left({1\over 2\kappa^2}\, R(g) - V(T)\sqrt{1 -
2\pi\ell_s^2 f(T) (\dot T)^2}\,\right),
\end{equation} 
where, the relation
\begin{equation}\label{planck}
{\sqrt{8\pi}\over m_{Pl}} = \kappa = {(2\pi)^2 g_s\ell_s\over\sqrt{2 v}}
\end{equation} 
expresses the inverse reduced Planck mass $\kappa$ in terms of
the string tension and string coupling $g_s$. In \refb{planck},
$v$ is a dimensionless parameter corresponding to the volume 
of the 22- or 6-dimensional space transverse to the 3-brane. 
It depends on how the four dimensional spacetime is realized 
from the 26- or 10-dimensional one. 
  
The Lagrangian \refb{gravT}, with $f(T)=1$, has recently 
been analyzed in 
Refs.\cite{Gibb,FaTy}. The inclusion of $f(T)$ leads to some
modifications. The Hamiltonian or energy density of $T$ 
is\footnote{We set $k=0$ for simplicity.}
\begin{equation}\label{hamil}
{\cal H} = {a^3(t) V(T)\over\sqrt{1- 2\pi\ell_s^2 f(T){\dot T}^2}} 
= a^3(t)\,\rho(T)
\end{equation} 
and its equation of motion is
\begin{equation}\label{roeom}
\dot\rho = - 3 H (p + \rho),
\end{equation} 
where, $H=\dot a/a$ is the Hubble parameter and 
$p=-V(T)\sqrt{1- 2\pi\ell_s^2 f(T){\dot T}^2}$ is the pressure 
of the tachyonic fluid. The explicit expressions of 
$p$ and $\rho$ may be substituted to obtain the second order 
evolution equation of $T$.

The gravitational equations, on the other hand, are
\begin{eqnarray}
H^2 &=& {\kappa^2\over 3}\,\rho \;=\;
{\kappa^2\over3}\,{V(T)\over\sqrt{1- 2\pi\ell_s^2 f(T){\dot T}^2}},
\label{FRone}\\
\dot H &=& -\,{\kappa^2\over2}\,\left(\rho + p\right)\;=\;
-{\kappa^2\over2}\,{V(T)f(T)\over\sqrt{1- 2\pi\ell_s^2 
f(T){\dot T}^2}}\,2\pi\ell_s^2 {\dot T}^2.\label{FRtwo}
\end{eqnarray} 
Eqn.\refb{FRone} may be rewritten as 
$\sqrt{2\pi\ell_s^2 f(T)}\;{\dot T} = \sqrt{\left(1 - 
\kappa^4 V^2(T)/9H^4\right)}$. 


\section{Inflation from tachyon on a bosonic D-brane}
The effective action of the tachyon field $T(x)$ determined 
in the framework of bosonic B-SFT is\cite{GeSh,KuMaMoB}
\begin{equation}\label{BbsftS}
S_B = \tau_3\int d^4x\,\left( \ell_s^2 e^{-T}\,\p_\mu T\p^\mu T 
+ (T+1)\,e^{-T}\right),
\end{equation} 
where the normalization factor 
\begin{equation}\label{btension}
\tau_3 = {1\over(2\pi\ell_s^2)^2 2\pi g_s} =
{\pi^3 g_s^3\over 8 v^2}\, m_{Pl}^4
\end{equation}  
is the tension of the D$3$-brane\cite{GhS}. In our convention, 
the tachyon field $T(x)$ is dimensionless. Hence 
$V(T)=\tau_3(T+1)\exp(-T)$ has mass dimension four. It is 
parametrized by the dimensionless quantities $v$ and $g_s$.

Upto two derivative terms, the action \refb{BbsftS}\ is exact, in the 
sense that it incorporates the effect of all open string modes. The
potential has a maximum corresponding to the D-brane at $T=0$ and
a minimum at $T=\infty$. (It is also unbounded for negative values 
of $T$, but that is a pathology of the bosonic theory.) Notice
the $e^{-T}$ factor due to which the kinetic term has a non-standard
form. Hence the distance between the maximum and minimum (in 
field space) is finite. 

It is of course possible to do a field redefinition to bring the
kinetic term to the canonical form $\p_\mu\phi\p^\mu\phi$.
The transformation is $\phi=2e^{-T/2}$, in terms of which
the Lagrangian in
\begin{equation}\label{philag}
{\cal L}_\phi = \tau_3\,\left(\ell_s^2 \p_\mu\phi\p^\mu\phi - 
{1\over4}\phi^2\left(\ln{\phi^2\over4} - 1\right)\right).
\end{equation} 
We will come back to the description of the dynamics in terms of the 
field $\phi$ later on in this section. 

There are an infinite number of corrections to the action 
\refb{BbsftS}. In particular, the Born-Infeld action \refb{BornInf}\ 
gives a good description of slowly varying fields, {\em i.e.}, those 
for which the second and higher derivatives of the field can be 
ignored. Let us now treat the action \refb{BbsftS}\ as an expansion 
of an action of the form \refb{BornInf}
\begin{equation}\label{BtachBI} 
S_{BI} = \tau_3\int d^4x\,(1+T)\, e^{-T}\left(
1+\pi\ell_s^2 f(T)\p_\mu T\p^\mu T\right).
\end{equation} 
{}From a comparison of eqns.\refb{BbsftS} and \refb{BtachBI}, in
a region where both are good descriptions, it follows that 
$f(T)= 1/(\pi (1+T))$. Put in another way, this factor is the
necessary field redefinition to relate the tachyon fields in
the Born-Infeld and B-SFT descriptions. 

Once again, it is possible to do 
a field redefinition to soak up the factor of $f(T)$, but this does 
not map the entire field space $T(x)\in(-\infty,\infty)$ to that of 
the redefined tachyon $\tilde T(x)= \sqrt{1+T(x)}$. Interestingly, it 
does, however, map the domain $-1\leq T(x) \leq \infty$, which is 
precisely the region accessed by all the classical solutions, 
{\em viz.}, the D-branes. Another amusing aspect of this field 
redefinition is that the potential in terms of $\tilde T(x)$ has a
Gaussian factor, reminiscent of the superstring case. It should,
however, be noted that the late time evolution of the tachyon field
in either of these two cases does not seem to match the asymptotic
behaviour analyzed by Sen\cite{senFT}. 
Of course, a field redefinition  could reproduce the correct 
asymptotics, but it would, 
necessarily, have to  involve derivatives of the tachyon field. 
Since we are mainly interested in the early time evolution
where the time derivatives of the tachyon turn out to be small, 
the aforementioned derivative terms do not play a significant role. 
Furthermore, a field redefinition of the tachyon would not change 
the dynamics of the scale factor of the universe. Hence we may 
conclude that the results we obtain are not very sensitive to the 
large time behaviour.

\begin{figure}[!ht]
\leavevmode
\vspace*{-2.5truein}
\begin{center}
\epsfxsize=4truein
\epsfbox{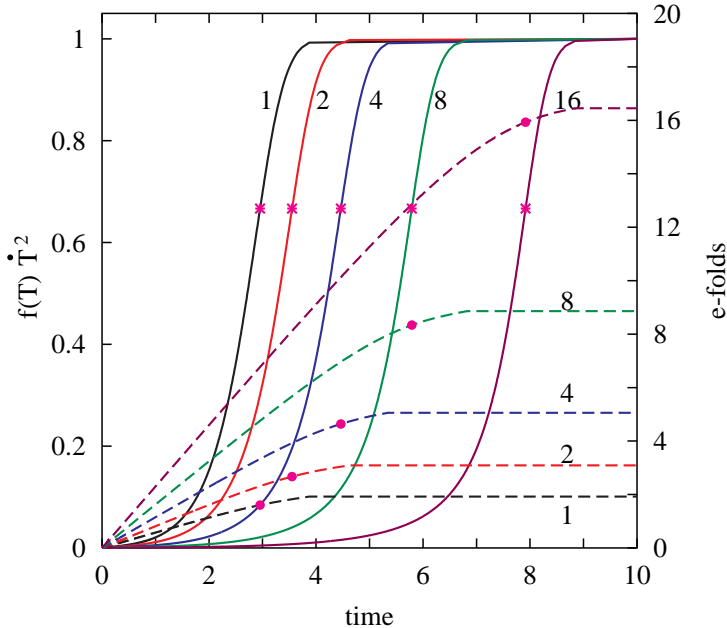}
\caption{The kinetic term $f(T) \dot T^2$ (solid lines) and 
e-foldings (dashed lines) vs time $t$ for bosonic string. Numbers
labelling the graphs correspond to values of $\kappa^2\tau_3$.
In each case, the point at which the inflation ends has been marked.} 
\label{bosfig}
\end{center}
\end{figure}

One can now solve the equations of motion \refb{FRone}\ and 
\refb{FRtwo}\ numerically to determine the evolution of the 
tachyon and the Hubble parameter. Following \cite{senroll}, we 
choose the initial conditions such that the tachyon is 
infinitesimally displaced to the right of the maximum of the 
potential\footnote{For the numerical computation, we set the string 
tension $1/2\pi\ell_s^2=1$. (The tension of the 3-brane $\tau_3$
is therefore $1/2\pi g_s$.) The actual seed used for the initial
value of the tachyon is $T\sim 0.1$, which is compatible with the 
requirement in\cite{LindeF}.} and has zero velocity. 
The results depend on the tension
$\tau_3$ of the D3-brane, and are displayed in 
Fig.\ref{bosfig}. In the same figure, we also plot the number 
of e-foldings
\begin{equation}\label{efold}
N_e(t)\; =\; \ln{a(t_f)\over a(t_i)} \;=\;\int_{t_i}^{t_f}dt\,H(t)
\end{equation} 
as a function of time. It is clear from the plot that the effective
kinetic term $2\pi\ell_s^2 f(T){\dot T}^2$ initially remains very 
close to zero, during which the expansion takes place. Subsequently,
the tachyon rolls very fast and the kinteic energy saturates to the 
maximum value of $2\pi\ell_s^2 f(T){\dot T}^2=1$. The exit from the 
inflationary phase occurs very soon during the fast roll at the point
$2\pi\ell_s^2 f(T){\dot T}^2=2/3$.
 
Finally, we would like to mention that in terms of the field 
$\phi=2e^{-T/2}$, which has a canonical kinetic term \refb{philag},
one obtains $f(\phi)=2/V(\phi)$. It is easy to check that one
gets the same amount of e-folding from the evolution of $\phi$.
This is but a reaffirmation of the fact that field redefinitions cannot
change the physical consequences of the dynamics.


\section{Inflation from tachyon on a non-BPS D-brane}
The analysis of the previous section can easily be extended to the 
case of the superstring. Recall, that the effective action of the 
tachyon field ${\cal T}(x)$ on a non-BPS D$3$-brane\footnote{One can 
also study the complex tachyon on a non-supersymmetric brane-antibrane
pair\cite{DDbar}. The qualitative features of this system, however, 
are expected to be the same as those of a non-BPS brane.} computed 
in supersymmetric B-SFT is\cite{KuMaMoS}
\begin{equation}\label{SbsftS}
S_F = \tau_3\int d^4x\,\left(\ell_s^2 \ln 2\,e^{-{\cal T}^2/4}\,
\p_\mu{\cal T}\p^\mu{\cal T} + e^{-{\cal T}^2/4}
\right),
\end{equation} 
where, $\tau_3$, the tension of the brane is the overall normalization
factor\cite{Ghsnorm}. As in the bosonic case, this action is exact
upto two derivatives. Using arguments identical to those in the previous 
section, we can write down the Born-Infeld form of the action
\begin{equation}\label{StachBI}
S_{BI} = \tau_3\int d^{4}x\,e^{-{\cal T}^2/4}\,\sqrt{1-
2\ell_s^2 \ln 2\,(\dot{\cal T})^2}.
\end{equation} 
In other words, $f({\cal T})=(\ln 2)/\pi$.

\begin{figure}[!ht]
\leavevmode
\vspace*{-2.5truein}
\begin{center}
\epsfxsize=4truein
\epsfbox{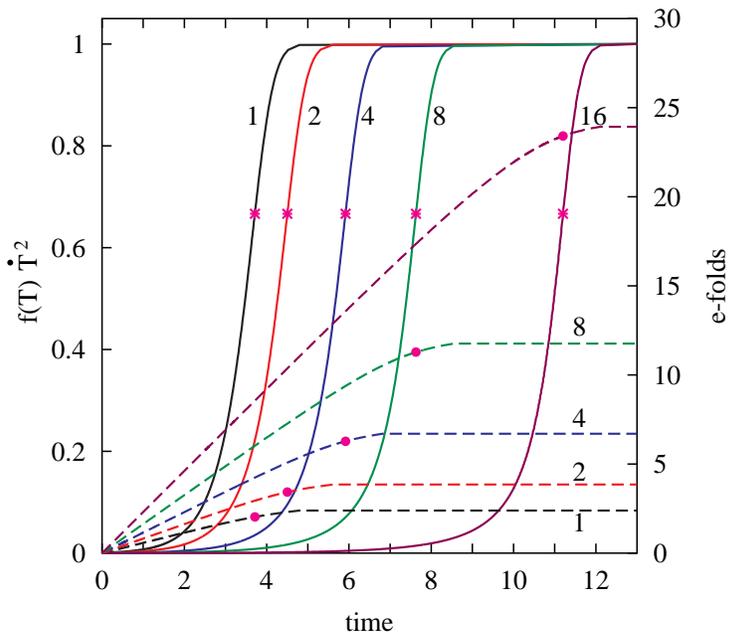}
\caption{The kinetic term $f(T) \dot T^2$ (solid lines) and 
e-foldings (dashed lines) vs time $t$ for the superstring. The 
parameters refer to $\kappa^2\tau_3$.} 
\label{superfig}
\end{center}
\end{figure}

As in the previous section, we now solve the equations of motion 
numerically to obtain the evolution of the tachyon and the Hubble 
parameter. Once again, the initial conditions are chosen such that 
the tachyon field starts with a small positive value and zero velocity. 
The results are displayed in Fig.\ref{superfig}. The qualitative
features are similar to those in the bosonic case, although 
for similar values of the brane tension $\tau_3$, the tachyon of
the superstring has small velocity for a somewhat longer time, 
which results in a few more e-foldings. 

\section{Remarks on tachyon driven inflation}
In order to generate enough inflation, it is necessary for the
inflaton field to roll slowly enough. This is characterized
by two dimensionless parameters $\varepsilon$ and $\eta$. For
the conventional inflaton with canonically normalized kinetic 
energy term, these are given by\cite{LL}
\begin{equation}\label{cslow}
\varepsilon = {m_{Pl}^2\over 2}\,\left({V'\over V}\right)^2,
\qquad\qquad
\eta = m_{Pl}^2\,\left({V''\over V}\right).
\end{equation} 
The conditions for slow roll inflation are $\varepsilon \ll 1$ and
$|\eta| \ll 1$.
These formulas are not directly applicable to case of the tachyon, 
which has a non-standard action \refb{BtachBI} or \refb{StachBI}.
Using the field redefinition discussed in Sec.3, it is easy to 
show that
\begin{eqnarray}
\varepsilon &=& {m_{Pl}^2\over 4\pi\ell_s^2f(T)V(T)}\,
\left({V'(T)\over V(T)}\right)^2,\nonumber\\
\eta &=& - \varepsilon + {m_{Pl}^2\over 2\pi\ell_s^2f(T)V(T)}\,
\left({V''(T)\over V(T)}-{f'(T)V'(T)\over 2f(T)V(T)}\right)
\label{tslow},
\end{eqnarray}  
are the appropriate parameters. 

\begin{figure}[!ht]
\leavevmode
\vspace*{-3truein}
\begin{center}
\epsfxsize=4.5truein
\epsfbox{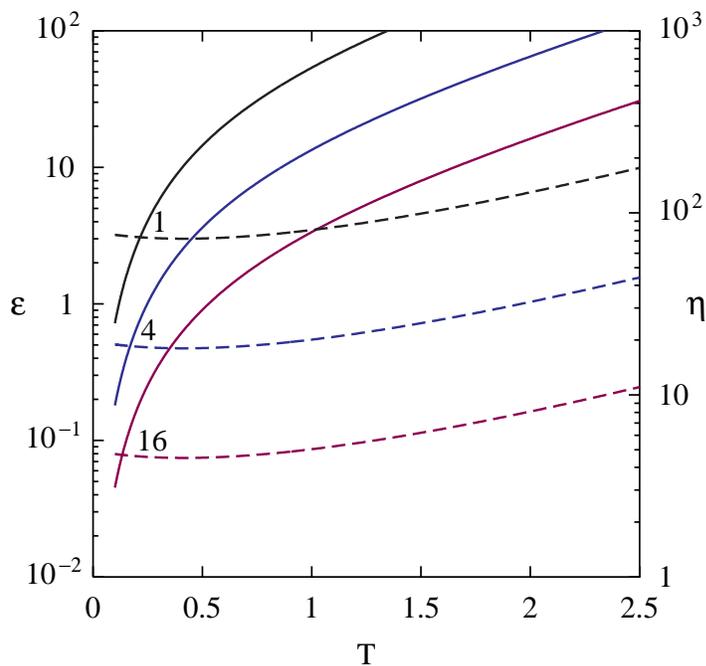}
\caption{Slow roll parameters $\varepsilon$ (solid lines) and 
$\eta$ (dashed lines) as a function of
tachyon for the bosonic theory. Numbers 
labelling the graphs correspond to values of $\kappa^2\tau_3$.}
\label{epseta}
\end{center}
\end{figure}

There is, nevertheless, a difference with conventional inflaton, 
especially if we are in the 
weak string coupling limit (small to moderate values 
of $\kappa^2 \tau_3$). For the potential dictated by 
either the bosonic string or the superstring theory, the  
consequent inflation is not of slow roll type\cite{koflin}. 
For example, while the slow roll parameter
$\varepsilon$ is small at the top of the potential, 
$\eta$ is not. In fact, as Fig.~\ref{epseta} demonstrates 
for the bosonic theory\footnote{It is easy to see that while 
  $\varepsilon$ for the superstring tachyon starts arbitrarily 
  close to zero, $\eta$, once again, is always large if one 
  is in the weak coupling limit.},
$\eta$ is significantly larger than unity
for the entire length of the evolution governed by tachyon 
dynamics. Thus, while the inflation satisfies 
one slow roll criterion (at least, initially), 
it fails to satisfy the other. The universe fails to inflate
enough as a consequence of this. 

\begin{figure}[!ht]
\leavevmode
\vspace*{-1.5truein}
\begin{center}
\epsfxsize=3truein
\epsfbox{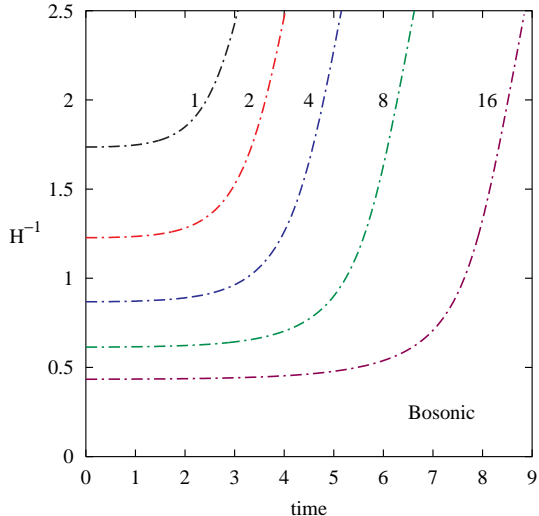}
\caption{The inverse Hubble parameter, in the bosonic theory,
as a function of time. Numbers 
labelling the graphs correspond to values of $\kappa^2\tau_3$.}
\label{hinverse}
\end{center}
\end{figure}

Now this may not be all bad news, for the situation here is 
similar to the fast roll inflation~\cite{LindeF}. 
It has been argued that such a fast roll inflation could have 
preceded the conventional slow roll inflation and could, in fact, 
have set the stage for the latter by suitably adjusting the
initial conditions for slow roll. In this sense, it is satisfying
to see that the inflation driven by the tachyon does not give rise 
to sufficient number of e-foldings, a job which will be relegated
to the slow roll inflation that follows. After we exit the early epoch
of tachyon driven fast roll inflation, the scale factor grows according
to a power law (see Fig.\ref{hinverse}). However, if a new scalar 
field takes over at this epoch and leads to slow roll inflation then 
the scale factor will continue to grow exponentially. Such details, 
though interesting, are beyond the scope of this work. Suffice is
to note that there are enough candidates for the scalar among the 
open string (transverse scalars on the brane) and closed string 
moduli. Let us note in passing that in slow roll inflation, the 
ratio of tensor to scalar amplitudes is proportional to $\epsilon$ 
and observations demand it to be of the order of $10^{-2}$ or smaller. 
Since in our case $\epsilon$ is of the order of $10^{-1}$, this ratio 
is larger than the observed 
value\footnote{L.\ Kofman, private communication.}. 
However, if the tachyon driven fast roll inflation is followed by 
a slow roll inflation then it is conceivable that it may readjust 
this ratio to the desired value. One can also imagine modification 
of the present scenario of inflation on a single brane by considering 
a stack of multiple branes as considered in \cite{early1,FaTy}. 
 
It is worth emphasizing another feature of inflation driven by
the tachyon field. The tachyon potential is not flat. Indeed,
the leading term in the effective tachyon potentials 
\refb{BbsftS} and \refb{SbsftS} from string theory is quadratic 
at the maximum (as befits a tachyon). The kinetic term of the 
tachyon field, on the other hand, is non-trivial. This leads to 
all the difference in their dynamics. 


\section{Quintessence vs rolling tachyon}
Recent observations indicate that there is probably an 
infinitesimally small positive vacuum energy density in the 
universe. This could be due to a
cosmological constant. An alternative to this idea is that of 
quintessence, a conventional scalar field $\chi$ with a potential 
$V(\chi)$, such that $\chi$ is still rolling towards its minimum 
(see \cite{suss}\ and references therein). Generically the 
minimum of the potential is taken to be at $\chi = \infty$. In 
this section, we examine whether the rolling tachyon could be 
an answer to this problem. 

The equation of state for quintessence is $p=\omega_\chi\rho$ with 
\begin{equation}\label{qomega}
-1 < \omega_\chi < -{1\over 3}.
\end{equation}
The evolution of a scalar field in a homogeneous and isotropic
universe described by the FRW metric \refb{FRW}\ (with $k=0$ for 
simplicity) is given by \refb{roeom}. This, along with the equation 
of state, determine the energy density $\rho$ in terms of the 
scale factor $a(t)$. Eqn.\refb{FRone}\ can now be solved to get
\begin{equation}\label{quintscale}
\rho = \rho_0 \left({a_0\over a}\right)^{3(1+\omega)},\qquad\qquad 
a(t) = a_0 \left( {t\over t_0}\right)^{{2\over 3(1+\omega)}}.
\end{equation}
It is clear from the power law behaviour of the scale factor that 
the Hubble parameter $H$ is inversely proportional to time, 
indicating a curvature singularity at $t=0$. This is because the
curvature tensors are proportional to $H$ and $\dot H$. As time 
evolves, curvature goes to zero and spacetime becomes flat. 

The equations of motion for the tachyon field, when expressed in 
terms of the energy density and pressure, are identical to that
of an ordinary scalar field. Its equation of state, however, is 
fundamentally different\cite{senroll}. The parameter $\omega$ is 
not a constant but instead is a time dependent function. More
specifically,  
\begin{equation}\label{tomega}
\omega(T) = 2\pi\ell_s^2 f(T) \dot T^2 - 1,
\end{equation}
as follows from the expressions of $\rho$ and $p$ in Sec.2.
With the initial conditions we have chosen in Secs. 3 and 4, namely
$T(0)=0+$ and $\dot T(0)=0$, $\omega$ starts with the value $-1$. 
It is easy to see that the energy density at this stage is independent 
of the scale factor $a(t)$. Substituting this into the Einstein's 
equation  \refb{FRone} we find that the scale factor grows 
exponentially,
\begin{equation}\label{qstart}
\rho = \rho_0, \qquad\qquad a(t) = a_0\,e^{t/t_0},
\end{equation}
with $\rho_0=\tau_3$ and arbitrary constants $a_0$ and $t_0$. 
This means that in the early epoch the scale factor grows 
exponentially giving a constant Hubble parameter $H$. Thus unlike 
the quintessence scalar field, tachyon dynamics does not lead to a 
curvature singularity as $t\to 0$. 

As the system evolves, $\omega(T)$ moves away from $-1$. We see
from the Figs.~\ref{bosfig} and \ref{superfig} that initially 
$2\pi\ell_s^2 f(T)\dot T^2 \ll 1$ for some time. This in turn means 
that $\omega\simeq -1$ (but $\omega\ne -1$), is virtually a constant 
during this epoch. The system evolves according to 
eqn.\refb{quintscale}, {\em i.e.}, the scale factor grows 
according to a power law with a
large exponent. Close to the end of the inflationary era, 
$f(T)\dot T^2$ grows rapidly, leading to an increase in the value
of $\omega$, which eventually settles down to $\omega = 0$. 

In summary, we see that the early time development of the tachyon is 
quite different from that of the conventional quintessence scalar field. 
In particular, there is no curvature singularity as $t\to 0$;
instead there is a constant scale factor and a flat metric. With 
the rolling of the tachyon, initial exponential growth of 
the scale factor switches over to a power law growth (see
Fig.\ref{hinverse}).
Eventually, as the rolling tachyon picks up speed, growth of $a(t)$ 
as a function of $t$ slows down and settles at $t^{2/3}$ indicating 
a transition to a tachyon (and other) matter dominated era. It will 
be interesting to investigate tachyon driven quintessence models 
in detail. We hope to come back to it in future. 

\vspace*{1ex}
\noindent{\bf Acknowledgement:} It is a pleasure to thank Lev Kofman
and Ashoke Sen for valuable comments and discussions. DC would like 
to thank the Department of Science and Technology, Government of 
India for financial assistance under Swarnajayanti Fellowship. 



\newpage

\begin{thebibliography}{99}

\bibitem{origin}
A.\ Sen, JHEP {\bf 9808} (1998) 010 [{\tt hep-th/9805019}];
JHEP {\bf 9808} (1998) 012 [{\tt hep-th/9805170}];
JHEP {\bf 9912} (1999) 027 [{\tt hep-th/9911116}].

\bibitem{as}
A.\ Sen, JHEP {\bf 9910} (1999) 008 [{\tt hep-th/9909062}];
J.\ Math.\ Phys.\  {\bf 42} (2001) 2844 [{\tt hep-th/0010240}].

\bibitem{senroll}
A.\ Sen, JHEP 0204 (2002) 048, [{\tt hep-th/0203211}].

\bibitem{senmatt}
A.\ Sen, {\em Tachyon matter}, {\tt hep-th/0203265}.

\bibitem{Gibb}
G.\ Gibbons, Phys.\ Lett.\ B {\bf 537} (2002) 1, 
[{\tt hep-th/0204008}].

\bibitem{FaTy}
M.\ Fairbairn and M.\ Tytgat, {\em Inflation from a tachyon fluid?},
{\tt hep-th/0204070}.

\bibitem{mukoh}
S.\ Mukohyama, {\em Brane cosmology driven by the rolling tachyon},
{\tt hep-th/0204084}.

\bibitem{senFT}
A.\ Sen, {\em Field theory of tachyon matter}, {\tt hep-th/0204143}.

\bibitem{paddy}
A.\ Feinstein, {\em Power-law inflation from the rolling tachyon}, {\tt
hep-th/0204140}; 
T.\ Padmanabhan, {\em Accelerated expansion of the universe driven by tachyon
matter}, {\tt hep-th/0204150}.

\bibitem{early}
C.\ Burgess, M.\ Majumdar, D.\ Nolte, F.\ Quevedo, G.\ Rajesh
and R.\ Zhang, JHEP 0107 (2001) 047 {\tt hep-th/0105204}.

\bibitem{early1}
A.\ Mazumdar, S.\ Panda and A.\ Perez-Lorenzana,
Nucl.\ Phys.\ B {\bf 614} (2001) 101 [{\tt hep-ph/0107058}].


\bibitem{effective}
M.\ Garousi, Nucl.\ Phys.\ B {\bf 584} (2000) 284 
[{\tt hep-th/0003122}];\\
E.\ Bergshoeff, M.\ de Roo, T.\ de Wit, E.\ Eyras and S.\ Panda,
JHEP {\bf 0005} (2000) 009 [{\tt hep-th/0003221}];\\
J.\ Kluson, Phys.\ Rev.\ D {\bf 62} (2000) 126003 (2000)
[{\tt hep-th/0004106}].

\bibitem{GeSh}
A.\ Gerasimov and S.\ Shatashvili, JHEP {\bf 0010} (2000) 034 
[{\tt hep-th/0009103}]. 

\bibitem{KuMaMoB}
D.\ Kutasov, M.\ Mari\~ no and G.\ Moore, JHEP {\bf 0010} (2000) 045
[{\tt hep-th/0009148}].

\bibitem{KuMaMoS}
D.\ Kutasov, M.\ Mari\~ no and G.\ Moore, {\em Remarks on tachyon 
condensation in superstring field theory}, [{\tt hep-th/0010108}].

\bibitem{bsft}
E.\ Witten, Phys.\ Rev.\ D {\bf 46} (1992) 5467 [{\tt 
hep-th/9208027}]; Phys.\ Rev.\ D {\bf 47} (1993) 3405
[{\tt hep-th/9210065}].

\bibitem{koflin}
L.\ Kofman, A.\ Linde, JHEP 0207 (2002) 004,
[{\tt hep-th/0205121}].

\bibitem{ShW}
G.\ Shiu and I.\ Wasserman, {\em Cosmological constraints on tachyon
matter}, {\tt hep-th/0205003}.

\bibitem{GhS}
D.\ Ghoshal and A.\ Sen, JHEP 0011 (2000) 021 
[{\tt hep-th/0009191}].

\bibitem{DDbar}
P.\ Kraus and  F.\ Larsen, Phys.\ Rev.\ D {\bf 63} (2001)
106004 [{\tt hep-th/0012198}];\\
T.\ Takayanagi, S.\ Terashima, T.\ Uesugi, JHEP 0103 (2001) 
019 [{\tt hep-th/0012210}].

\bibitem{Ghsnorm}
D.\ Ghoshal, {\em Normalization of the Boundary Superstring Field Theory 
Action}, {\tt hep-th/0106231}.

\bibitem{LL}
A.\ Liddle and D.\ Lyth, {\em Cosmological inflation and large scale
structure}, CUP (2000). 

\bibitem{LindeF}
A.\ Linde, JHEP 0111 (2001) 052, [{\tt hep-th/0110195}].

\bibitem{suss}
S.\ Hellerman, L.\ Susskind and N.\ Kaloper, JHEP 0106 (2001) 
003 [{\tt hep-th/0104180}].

\end{thebibliography}
\end{document}